\documentclass[aps,pra,reprint]{revtex4-2}
\usepackage{graphicx} 
\usepackage{amsmath,amssymb,amsfonts}
\usepackage{url}

\usepackage[colorlinks=true, urlcolor=blue,linkcolor=blue,citecolor=blue]{hyperref}

\def\twocolumnbreak{\\}

\begin{document}

\title{Characterization and virtualization of a medical ultrasound transducer}

\author{Nathan Blanken}
\author{Michel Versluis}
\author{Guillaume Lajoinie}
\affiliation{Physics of Fluids Group, Technical Medical (TechMed) Centre, University of Twente, Enschede, The Netherlands}

\begin{abstract}
    With the rise of data-driven ultrasound imaging technologies, realistic simulation of ultrasound fields and radio-frequency data is becoming increasingly important.
    Accurate transducer characterization is crucial for realistic simulations.
    In this work, we present a streamlined ultrasound characterization pipeline that creates a virtual transducer model from acoustic holography measurements of the ultrasound field.
    The pipeline enables the extraction of the transmit impulse response, the receive impulse response, the size of the transducer aperture, and the focal distance of the lens.
    The method relies on acoustic field projections using either the angular spectrum method or Rayleigh integrals.
    Additionally, we present a method to compensate for misalignment in the measurement setup, based on the angular spectrum of the holographic measurement.
    We demonstrate the application of the characterization pipeline to a P4-1 phased array transducer.
    The resulting virtual transducer model can be imported into the PROTEUS simulation software for the generation of physically realistic ultrasound fields and ultrasound radio-frequency data with arbitrary transmit settings.
    The characterization pipeline has been released as an open-source toolkit to enable the ultrasound community to perform transducer characterization with increased accuracy and efficiency.
\end{abstract}

\maketitle


\section{Introduction}

Transducer characterization plays a key role in the field of medical ultrasound imaging.
Firstly, acoustic output measurements are required to ensure the safe implementation of new imaging systems and strategies~\cite{FDA2023}.
Furthermore, an accurate description of the transmit impulse response, the receive impulse response, and the beam profile of a transducer is essential for model-based image reconstruction algorithms and for accurate numerical simulation of ultrasound fields.
For example, the standard delay-and-sum reconstruction requires an accurate estimate of the pulse duration, the aperture size, and the focal distance~\cite{Perrot2021}.
Super-resolution strategies such as ultrasound localization microscopy require an accurate estimate of the point spread function of a microbubble point scatterer~\cite{Heiles2022}, which is closely related to the transmit and receive impulse responses of the transducer.
Recently, deep-learning based (super-resolution) ultrasound imaging strategies have been gaining increasing attention~\cite{Youn2020,Sloun2021,Blanken2022}.
These methods rely on physically realistic synthetic (radio-frequency) data to train neural networks.
Such simulations can be performed with ultrasound simulators such as k-Wave~\cite{Treeby2010} or PROTEUS~\cite{Blanken2024} in conjunction with an accurate model of the medical ultrasound transducer.
Moreover, accurate numerical models of a transducer allow for a rapid prediction of the acoustic field whenever experiments with new transmit settings are performed, eliminating the need for new acoustic output measurements, which are time-consuming.

Transducer characterization can be as simple as measuring the two-way response by placing a reflector in front of the transducer or measuring the acoustic pressure output with a microphone positioned in front of the transducer~\cite{Chen1997,Neer2010,Blanken2024}.
However, accurate implementation of such methods requires a good estimate of the spatial impulse response of the system~\cite{Jensen1992,Emeterio1992}, as the pressure field is spatially dependent.
To estimate the spatial impulse response, prior knowledge of the aperture size and the focal distance is required.
Multiple works have focused on predicting the response of a transducer with equivalent circuit models, such as the KLM (Krimholtz, Leedom, Matthaei) model~\cite{Leedom1971,Krimholtz1972,Castillo2003}.
However, these models require detailed knowledge of the piezoelectric oscillators and the internal architecture of the transducer, making them less suitable for commercial medical transducers, as this information may not be readily available.

Acoustic holography allows for the reconstruction of the three-dimensional acoustic field from a two-dimensional measurement plane and thus also allows for the vibration of the source to be reconstructed~\cite{Lee1985,Clement2000,Kreider2013}.
Unlike optical holography, acoustic holography does not require a reference beam to determine the phase of the wave, as the phase can be directly measured using a hydrophone.
Sapozhnikov~\textit{et al.} give an excellent overview of ultrasound characterization with acoustic holography~\cite{Sapozhnikov2015}, including a detailed error analysis.
Nonetheless, many works on acoustic holography only consider continuous wave emission and do not characterize the transient response of the source.
Nor do they describe how the reconstructed source can be converted into a numerical model for acoustic simulation software such as k-Wave~\cite{Treeby2010}.
The latter point is addressed by Treeby~\textit{et al.}, who use a gradient descent optimization scheme to find an equivalent source representation that can be use to simulate acoustic fields in k-Wave~\cite{Treeby2018}.
However, this approach is computationally expensive.
Moreover, many studies do not extract system parameters from the data such as the focal distance, and do not quantify the receive characteristics of the transducer.

In this work, we present a complete acoustic characterization pipeline from acoustic holography measurements to a virtual transducer model that can be used in k-Wave or PROTEUS to simulate ultrasound fields with arbitrary transmit properties.
We characterize both the transmit impulse response and the receive impulse response using field projections.
We provide a fitting procedure to find the aperture size and the focal distance of the transducer.
Additionally, we introduce a method to compensate for misalignment in the measurement setup based on the angular spectrum of the holographic measurement.
We demonstrate the pipeline by characterizing a P4-1 phased array.

In Section~\ref{sec:theory-methods}, we briefly review the diffraction theory required for transducer characterization, including the angular spectrum method and the Rayleigh integrals.
These projection methods use a Dirichlet boundary condition source representation, whereas Fourier collocation methods such as k-Wave typically use an embedded source representation.
We derive an analytical relation between these representations for linear, homogeneous, lossless media.
Next, we describe how the misalignment of the measurement plane with respect to the transducer can be extracted from the angular spectrum of the measurement data.
We conclude Section~\ref{sec:theory-methods} with an overview of the data processing pipeline.
In Section~\ref{sec:results} we demonstrate the characterization of a P4-1 transducer,  a 96-element 2.5-MHz phased array.
In addition to extracting the aforementioned system characteristics, we find that this P4-1 transducer also exhibits a strong spurious mode of oscillation.

Recognizing the need in the ultrasound community for a streamlined and easily accessible characterization procedure, we have translated the methodology presented in this paper into an open-source toolbox, available at (\url{https://github.com/PROTEUS-SIM/transducer-characterization})~\cite{Blanken2025github}.
The experimental dataset is available on Zenodo~\cite{Blanken2025zenodo}.
An earlier version of this text was published as part of Nathan Blanken's PhD thesis~\cite{Blanken2024thesis}.


\section{Theory and methods}\label{sec:theory-methods}

We define a virtual transducer as a holographic model~\cite{Sapozhnikov2015} of a transducer that can predict the full three-dimensional pressure field for any given voltage input $E_\mathrm{T}(t)$ and that can predict the recorded voltage output $E_\mathrm{R}(t)$ for an incoming pressure wave.
Throughout this work we will assume that a transducer is a linear, time-invariant system, such that the transducer can be characterized by finding a transmit impulse response $h_\mathrm{T}(t)$ and a receive impulse response $h_\mathrm{R}(t)$.
The normal surface velocity $u_\mathrm{T}(t)$ of the transducer can then be computed with
\begin{equation}\label{eq:transmit-impulse-response}
    u_\mathrm{T}(t) = h_\mathrm{T}(t)\ast E_\mathrm{T}(t),
\end{equation}
where $\ast$ denotes temporal convolution.
Alternatively, we could have defined the transmit impulse response in terms of the pressure at the transducer surface, rather than the velocity, a topic that we will come back to in Sections~\ref{sec:rayleigh-integrals} and~\ref{sec:source-discussion}.
Similarly, the receive response of the transducer can be computed with
\begin{equation}\label{eq:receive-impulse-response}
    E_\mathrm{R}(t) = h_\mathrm{R}(t)\ast p_\mathrm{R}(t),
\end{equation}
where $p_\mathrm{R}(t)$ is the pressure at the transducer surface.
An ultrasound transducer is often equipped with an acoustic lens, which has the effect of delaying the output signals across the transducer surface.
We therefore also aim to find the focal distance $F$ with which these delays can be computed.
Additionally, we aim to extract the size of the active transducer aperture, as it may not be known a priori.

To find the transmit velocity $u_\mathrm{T}(t)$ and the receive pressure $p_\mathrm{R}(t)$, we will perform a holographic measurement~\cite{Sapozhnikov2015} of the acoustic field.
We will briefly review two frequently used methods for projecting the measured field onto the transducer surface: the angular spectrum method and Rayleigh integrals.
These methods are most conveniently described in the frequency domain.
In much of the existing literature, a harmonic field $P(\mathbf{r};\omega)$ is understood to exhibit a time dependence $e^{-i\omega t}$.
To maintain this convention, we convert a time domain field $p(\mathbf{r},t)$ to a frequency domain field with the transformation
\begin{equation}\label{eq:complex-conjugate-fourier}
P(\mathbf{r};\omega) = \left(\mathcal{F}\{p(\mathbf{r},t)\}\right)^*,
\end{equation}
where $\mathcal{F}$ denotes the Fourier transform with respect to time, $i = \sqrt{-1}$, and the star denotes the complex conjugate.
The advantage of this notation is that plane waves are expressed as $e^{i\mathbf{k}\cdot\mathbf{r}-i\omega t}$, which ensures that they are travelling in the direction of the wave vector $\mathbf{k}$.
Similarly, we define the frequency domain of the normal particle velocity as
\begin{equation}
V(\mathbf{r};\omega) = \left(\mathcal{F}\{u_T(\mathbf{r},t)\}\right)^*
\end{equation}

We intend to construct a virtual transducer that can be incorporated in PROTEUS~\cite{Blanken2024}, which employs the toolbox k-Wave~\cite{Treeby2010} for its acoustic field computations.
However, in k-Wave energy is introduced into the system through embedded sources, rather than through Dirichlet boundary conditions.
We will therefore also derive relations between the holographic boundary conditions and the grid-based embedded source definitions.

In this work, we will use a Cartesian coordinate system ($x,y,z$) with its origin positioned at the centroid of the virtual transducer.
The coordinate $x$ is the coordinate in the lateral direction (across the transducer elements), $y$ is the coordinate in the elevation direction (along the transducer elements), and $z$ is the coordinate along the primary acoustic propagation direction (perpendicular to the transducer surface).

\subsection{Angular spectrum method}\label{sec:angular-spectrum}

The angular spectrum method has frequently been used to propagate acoustic pressure fields defined on a plane to another, parallel plane~\cite{Stepanishen1970,Williams1982,Kreider2013}.
For ease of reference, we will briefly outline the method here.
The angular spectrum of a harmonic pressure field is defined as the two-dimensional Fourier transform taken along a plane of interest at depth $z$~\cite{Stepanishen1982,Williams1982}:
\begin{equation}\label{eq:angular-spectrum}
    \hat{P}(k_x,k_y,z) = \int_{-\infty}^{\infty}\int_{-\infty}^{\infty}P(x,y,z)e^{-ik_xx - ik_yy}dxdy.
\end{equation}
The angular spectrum for the particle velocity normal to the plane is defined as:
\begin{equation}\label{eq:angular-spectrum-velocity}
    \hat{V}(k_x,k_y,z) = \int_{-\infty}^{\infty}\int_{-\infty}^{\infty}V(x,y,z)e^{-ik_xx - ik_yy}dxdy,
\end{equation}
The inverse operations are
\begin{multlinetwocolumn}\label{eq:angular-spectrum-inverse}
    P(x,y,z) = \twocolumnbreak \frac{1}{(2\pi)^2}\int_{-\infty}^{\infty}\int_{-\infty}^{\infty}\hat{P}(k_x,k_y,z)e^{ik_xx + ik_yy}dk_xdk_y,
\end{multlinetwocolumn}
and
\begin{multlinetwocolumn}\label{eq:angular-spectrum-velocity-inverse}
    V(x,y,z) = \twocolumnbreak \frac{1}{(2\pi)^2}\int_{-\infty}^{\infty}\int_{-\infty}^{\infty}\hat{V}(k_x,k_y,z)e^{ik_xx + ik_yy}dk_xdk_y,
\end{multlinetwocolumn}
The spectral propagators $G$ that relate the angular spectrum at $z=0$ to the angular spectrum at $z=d$ can be derived by inserting~\eqref{eq:angular-spectrum-inverse} into the Helmholtz equation (the time-independent form of the wave equation):
\begin{equation}\label{eq:helmholtz}
    \nabla^2P + k^2P = 0,
\end{equation}
where the wavenumber $k$ is defined as $k = \omega/c_0$, with $c_0$ the speed of sound in the medium.
For the relation between $\hat{P}$ and $\hat{V}$, we can make use of the momentum equation:
\begin{equation}
    \rho_0\frac{\partial \mathbf{u}}{\partial t} = -\nabla p.
\end{equation}
For the normal velocity along $z$ in the frequency domain, the momentum equation can be written as
\begin{equation}\label{eq:momentum-normal}
    i\rho_0c_0kV = \frac{\partial P}{\partial z}.
\end{equation}
Combining~\eqref{eq:angular-spectrum-inverse},~\eqref{eq:angular-spectrum-velocity-inverse},~\eqref{eq:helmholtz} and~\eqref{eq:momentum-normal} yields expressions for the spectral propagators $G$ (see~\cite{Stepanishen1982} for further details):
\begin{multlinetwocolumn}\label{eq:spectral-propagator-vp}
    \hat{P}(k_x,k_y,d) = \rho_0c_0G_\mathrm{vp}(k_x,k_y,d)\hat{V}(k_x,k_y,0) \twocolumnbreak
        \mathrm{~with~} G_\mathrm{vp} = \frac{k}{k_z}e^{ik_zd}
\end{multlinetwocolumn}
\begin{multlinetwocolumn}
    \hat{P}(k_x,k_y,d) = G_\mathrm{pp}(k_x,k_y,d)\hat{P}(k_x,k_y,0) \twocolumnbreak
        \mathrm{~with~} G_\mathrm{pp} = e^{ik_zd}
\end{multlinetwocolumn}
\begin{multlinetwocolumn}
   \hat{V}(k_x,k_y,d) = \frac{1}{\rho_0c_0}G_\mathrm{pv}(k_x,k_y,d)\hat{P}(k_x,k_y,0) \twocolumnbreak
        \mathrm{~with~} G_\mathrm{pv} = \frac{k_z}{k}e^{ik_zd}
\end{multlinetwocolumn}
and $k_z = \sqrt{k^2 - k_x^2 - k_y^2}$.
Since the definition of $P(\mathbf{r};\omega)$ in~\eqref{eq:complex-conjugate-fourier} also includes negative frequencies $\omega$, we also need a definition for $G(k_x,k_y,d;-k)$.
We define $G(k_x,k_y,d;-k) = G(k_x,k_y,d;k)^*$ to ensure that the time domain signals remain real.

\begin{figure*}[htp]
    \centering
    \includegraphics[width=13cm]{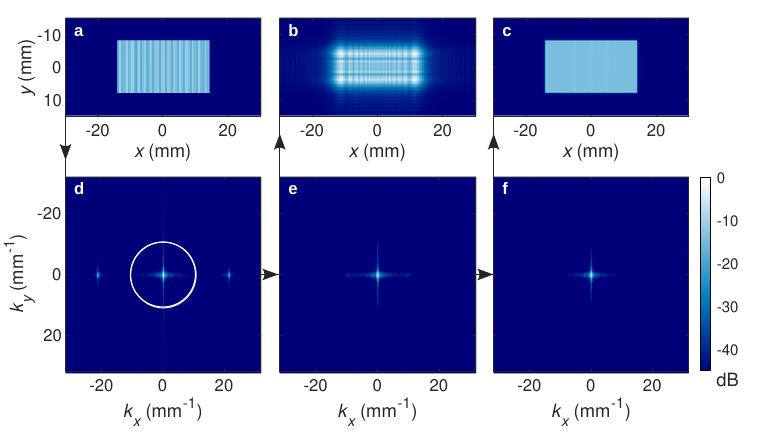}
    \caption{
        \textbf{Demonstration of the angular spectrum method.}
        \textbf{a} Rectangular aperture linear array.
        \textbf{b} Field at 20~mm from the source.
        \textbf{c} Reconstructed source.
        \textbf{d} Angular spectrum of the source.
        The white circle represents $k_x^2+k_y^2=k^2$.
        \textbf{e} Angular spectrum of the field at 20~mm.
        \textbf{f} Angular spectrum spectrum of the reconstructed source.
    }
    \label{fig:angular-spectrum}
\end{figure*}

As Williams and Maynard~\cite{Williams1982} point out, the spectral propagators contain rapid oscillations in $k$-space as $k_x^2 + k_y^2 \rightarrow k^2$.
Furthermore, \eqref{eq:spectral-propagator-vp} contains a singularity at $k_x^2 + k_y^2 = k^2$.
This can lead to large undersampling errors in discrete numerical implementation.
One way to alleviate this, is angular restriction, i.e. restricting the angular spectrum to $k$-vectors that are sufficiently far from the circle $k_x^2 + k_y^2 = k^2$~\cite{Zeng2008}.
Another, less restrictive method, was proposed by Williams and Maynard~\cite{Williams1982}, who note that, in the discrete evaluation, it is the average value of the spectral propagator around the sampling point in $k$-space that matters, which is a finite.
They evaluate the averaged spectral propagator in~\eqref{eq:spectral-propagator-vp} at $d=0$.
We have extended their evaluation to all three spectral propagators and general $d$ in the appendix.

Figure~\ref{fig:angular-spectrum} demonstrates the application of the angular spectrum method.
Figure~\ref{fig:angular-spectrum}a shows the transducer surface of a 96-element 2.5-MHz linear array.
Figure~\ref{fig:angular-spectrum}d shows its angular spectrum.
The white circle represents $k_x^2+k_y^2=k^2$, with $k$ the wavenumber corresponding to a frequency of 2.5~MHz in water.
The bright features outside the circle correspond to the rapidly oscillating pattern of the individual transducer elements.
As $\exp(ik_zd) = \exp(-\sqrt{k_x^2+k_y^2-k^2}d)$, these components only result in evanescent waves and are not propagated into the far field.
Figure~\ref{fig:angular-spectrum}e is obtained by applying the spectral propagator $G_\mathrm{vp}$ for $d=20$~mm.
Subsequently applying the spectral propagator $G_\mathrm{pv}$ for $d=-20$~mm reconstructs the source.
The reconstructed source shows a good correspondence with the ground truth source, except for the high-frequency spatial components that rapidly decay as evanescent waves.

\subsection{Rayleigh integrals}\label{sec:rayleigh-integrals}

The space domain equivalent of the spectral propagators are the Rayleigh integrals~\cite{Sapozhnikov2015,Williams1982,Bouwkamp1954}.
Sapozhnikov~\textit{et al.} list four versions of the Rayleigh integral, which we will reproduce here for ease of reference~\cite{Sapozhnikov2015}.
Forward wave propagation from a source surface $\Sigma_1$ to a sensor surface $\Sigma_2$ can be computed with
\begin{multlinetwocolumn}\label{eq:rayleigh-fwd-vp}
    P(\mathbf{r}_2) = \int_{\Sigma_1}V(\mathbf{r}_1)K_\mathrm{vp}^\mathrm{fwd}(\mathbf{r}_1;\mathbf{r}_2)d\Sigma_1 \\
    \mathrm{~with~} K_\mathrm{vp}^\mathrm{fwd}(\mathbf{r}_1;\mathbf{r}_2) =
    -\frac{ik\rho_0c_0}{2\pi}\frac{e^{ikR}}{R}
\end{multlinetwocolumn}
or with
\begin{multline}\label{eq:rayleigh-fwd-pp}
    P(\mathbf{r}_2) = \int_{\Sigma_1}P(\mathbf{r}_1)K_\mathrm{pp}^\mathrm{fwd}(\mathbf{r}_1;\mathbf{r}_2)d\Sigma_1 \\
    \mathrm{~with~} K_\mathrm{pp}^\mathrm{fwd}(\mathbf{r}_1;\mathbf{r}_2) =
    \frac{1}{2\pi}(\mathbf{m}_{12}\cdot\mathbf{n}_1)
    \left(\frac{-ik}{R}+\frac{1}{R^2}\right)e^{ikR}.
\end{multline}
Backward wave propagation from a sensor surface $\Sigma_2$ to a source surface $\Sigma_1$ can be computed with
\begin{multline}\label{eq:rayleigh-bwd-pp}
    P(\mathbf{r}_1) = \int_{\Sigma_2}P(\mathbf{r}_2)K_\mathrm{pp}^\mathrm{bwd}(\mathbf{r}_2;\mathbf{r}_1)d\Sigma_2 \\
    \mathrm{~with~} K_\mathrm{pp}^\mathrm{bwd}(\mathbf{r}_2;\mathbf{r}_1) =
    \frac{1}{2\pi}(\mathbf{m}_{21}\cdot\mathbf{n}_2)
    \left(\frac{ik}{R}+\frac{1}{R^2}\right)e^{-ikR}
\end{multline}
or with
\begin{multline}\label{eq:rayleigh-bwd-pv}
    V(\mathbf{r}_1) = \int_{\Sigma_2}P(\mathbf{r}_2)K_\mathrm{pv}^\mathrm{bwd}(\mathbf{r}_2;\mathbf{r}_1)d\Sigma_2 \\
    \mathrm{~with~} K_\mathrm{pv}^\mathrm{bwd}(\mathbf{r}_2;\mathbf{r}_1) = 
    \frac{1}{2\pi ik\rho_0c_0}\left\{(\mathbf{n}_{1}\cdot\mathbf{n}_2)\left(\frac{ik}{R} + \frac{1}{R^2}\right) \right. \\
    + \left.(\mathbf{m}_{12}\cdot\mathbf{n}_1)(\mathbf{m}_{21}\cdot\mathbf{n}_2)
    \left(\frac{3ik}{R} + \frac{3}{R^2} - k^2\right)\right\} \frac{e^{-ikR}}{R}.
\end{multline}
Here, $R = |\mathbf{r}_1 - \mathbf{r}_2|$, $\mathbf{n}_1$ is the normal to the source plane, pointing in the forward propagation direction, $\mathbf{n}_2$ is the normal to the sensor plane, pointing in the backward propagation direction, $\mathbf{m}_{12} = (\mathbf{r}_2-\mathbf{r}_1)/R$, and $\mathbf{m}_{21} = - \mathbf{m}_{12}$.

The forward Rayleigh integrals are exact solutions to the wave equation for a planar source surface~\cite{Bouwkamp1954}, and they are an accurate approximation for curved surfaces provided the radius of curvature is much larger than the wavelength~\cite{Sapozhnikov2015}.
In the absence of evanescent waves, the backward integrals are also exact solutions for planar surfaces~\cite{Shewell68}.
These four versions of the Rayleigh integral all derive from the Kirchhoff-Helmholtz integral~\cite{Bouwkamp1954}.
Therefore,~\eqref{eq:rayleigh-fwd-vp} should give the same result as~\eqref{eq:rayleigh-fwd-pp}, and~\eqref{eq:rayleigh-bwd-pp} should give the same results as~\eqref{eq:rayleigh-bwd-pv}.
Nevertheless, these different formulations have different interpretations.
The equations with $V(\mathbf{r}_1)$ are practical if the normal velocity is zero outside the acoustic aperture.
In this case, $\Sigma_1$ can be restricted to the aperture area, a condition also referred to as rigid baffle~\cite{Emeterio1992,Franco2011}.
Similarly, if the pressure is zero outside the aperture, the equations with $P(\mathbf{r}_1)$ are more practical, a condition also referred to as soft baffle.

Another interpretation is that the convolution kernel $K_\mathrm{vp}^\mathrm{fwd}$ is the expression for a monopole field, while the convolution kernel $K_\mathrm{pp}^\mathrm{fwd}$ is the expression for a dipole field.
The source can thus be considered to consist of either infinitesimal monopole sources or infinitesimal dipole sources.
This interpretation will be relevant for Section~\ref{sec:equivalent-source}.

The main advantage of the angular spectrum method over the Rayleigh integrals is that the angular spectrum method can be evaluated numerically much more rapidly, as it can leverage fast Fourier transform algorithms.
However, the angular spectrum method only works for parallel source and sensor planes, although the method can be extended to isosurfaces in curvilinear coordinate systems~\cite{Rehman2022}.
The Rayleigh integral is computationally heavier, but can be applied to inclined and moderately curved surfaces.
In principle, the angular spectrum can be propagated to a range of depths to reconstruct the three-dimensional field surrounding the inclined or curved surface of interest.
The pressure or velocity at the surface can then be obtained through interpolation.
However, measurement data will typically be Nyquist sampled, requiring three-dimensional sinc interpolation, which is also computationally expensive.

\subsection{Equivalent source representations}\label{sec:equivalent-source}

In k-Wave~\cite{Treeby2010}, energy is introduced into the system through embedded mass and force sources.
Although Dirichlet boundary conditions can be enforced in k-Wave, it is not trivial to formulate boundary conditions that are accurate, stable, and retain the efficiency of the computational method~\cite{Treeby2018}.
Here, we derive a relation between the embedded sources used in k-Wave and the Dirichlet boundary condition used in the Rayleigh integral.
The mass source term $S_\mathrm{M}$ in k-Wave is defined as a source term in the continuity equation~\cite{Treeby2020}:
\begin{equation}
    \frac{\partial \rho}{\partial t} = -\rho_0\nabla\cdot\mathbf{u} + S_\mathrm{M},
\end{equation}
where $\rho$ is the local density and $\mathbf{u}$ is the local particle velocity.
The force source term is defined as a source term in the momentum equation:
\begin{equation}
    \frac{\partial \mathbf{u}}{\partial t} = -\frac{1}{\rho_0}\nabla p + \mathbf{S}_\mathrm{F}
\end{equation}
Using these definitions and the pressure-density relation $p = c_0^2\rho$, the second-order wave equation becomes~\cite{Treeby2020}:
\begin{multlinetwocolumn}
    \nabla^2p - \frac{1}{c_0^2}\frac{\partial^2 p}{\partial^2 t} = -S(\mathbf{r},t), \twocolumnbreak
        \mathrm{~with~}
    S(\mathbf{r},t) = -\rho_0\nabla\cdot\mathbf{S}_F + \frac{\partial}{\partial t}S_\mathrm{M}
\end{multlinetwocolumn}

For time-harmonic sources, $S(\mathbf{r},t) = \bar{S}(\mathbf{r})e^{-i\omega t}$, the solution to this equation can be found using the Green's function $e^{ikR}/(4\pi R)$:
\begin{equation}
    P(\mathbf{r}_2) = \int_{\mathbb{R}^3}\bar{S}(\mathbf{r}_1)\frac{e^{ikR}}{4\pi R}d^3r_1,
\end{equation}
where $d^3r_1$ denotes the volume integration element.
For a mass source distribution, we have
\begin{equation}\label{eq:green-solution-mass}
    P(\mathbf{r}_2) = -\frac{i\omega}{4\pi}\int_{\mathbb{R}^3}\bar{S}_\mathrm{M}(\mathbf{r}_1)\frac{e^{ikR}}{R}d^3r_1.
\end{equation}
For a force source distribution, we have
\begin{multline}
    P(\mathbf{r}_2) = -\rho_0\int_{\mathbb{R}^3}\nabla\cdot\mathbf{\bar{S}}_\mathrm{F}(\mathbf{r}_1)\frac{e^{ikR}}{4\pi R}d^3r_1 = \\
    -\rho_0\int_{\mathbb{R}^3}\nabla\cdot\left(\mathbf{\bar{S}}_\mathrm{F}(\mathbf{r}_1)\frac{e^{ikR}}{4\pi R}\right)d^3r_1 \twocolumnbreak
    +\rho_0\int_{\mathbb{R}^3}\mathbf{\bar{S}}_\mathrm{F}(\mathbf{r}_1)\cdot\nabla\left(\frac{e^{ikR}}{4\pi R}\right)d^3r_1,
\end{multline}
where $\nabla$ operates on $\mathbf{r}_1$. If $\mathbf{\bar{S}_\mathrm{F}}(\mathbf{r})$ is continuously differentiable and has compact support, then it follows from the divergence theorem that the first term vanishes:
\begin{multline}\label{eq:green-solution-force}
    P(\mathbf{r}_2) = \frac{\rho_0}{4\pi}\int_{\mathbb{R}^3}\mathbf{\bar{S}}_\mathrm{F}(\mathbf{r}_1)\cdot\nabla\left(\frac{e^{ikR}}{R}\right)d^3r_1, = \\
    \frac{\rho_0}{4\pi}\int_{\mathbb{R}^3}\mathbf{\bar{S}}_\mathrm{F}(\mathbf{r}_1)\cdot\mathbf{m}_{12}\left(-ik + \frac{1}{R}\right)\frac{e^{ikR}}{R}d^3r_1,
\end{multline}
where $\mathbf{m}_{12} = (\mathbf{r}_2-\mathbf{r}_1)/R$.

Now, to find the equivalent source representations, we turn~\eqref{eq:rayleigh-fwd-vp} and ~\eqref{eq:rayleigh-fwd-pp} into volume integrals using a delta function notation:
\begin{equation}\label{eq:rayleigh-fwd-vp-volume}
    P(\mathbf{r}_2) = -\frac{ik\rho_0c_0}{2\pi}\int_{\mathbb{R}^3
    }V(\mathbf{r}_1)\frac{e^{ikR}}{R}\delta(n_1)d\Sigma_1dn_1
\end{equation}
and
\begin{multlinetwocolumn}\label{eq:rayleigh-fwd-pp-volume}
    P(\mathbf{r}_2) = \twocolumnbreak \frac{1}{2\pi}\int_{\mathbb{R}^3}P(\mathbf{r}_1)(\mathbf{m}_{12}\cdot\mathbf{n}_1)
    \left(-ik+\frac{1}{R}\right)\frac{e^{ikR}}{R}\delta(n_1)d\Sigma_1dn_1,
\end{multlinetwocolumn}
where $n_1$ is a curvilinear coordinate defined such that $\partial\mathbf{r}/\partial n_1 = \mathbf{n}_1$ at the transducer surface. Equating~\eqref{eq:green-solution-mass} and~\eqref{eq:rayleigh-fwd-vp-volume}, equating ~\eqref{eq:green-solution-force} and~\eqref{eq:rayleigh-fwd-pp-volume}, and reverting to the time domain gives the following expressions for the equivalent source:
\begin{equation}
    S_\mathrm{M}(\mathbf{r},t) = 2\rho_0u_n(\mathbf{r},t)\delta(n_1)
\end{equation}
\begin{equation}
    \mathbf{S}_\mathrm{F}(\mathbf{r},t) = \frac{2p(\mathbf{r},t)}{\rho_0}\delta(n_1)\mathbf{n_1}
\end{equation}

\begin{figure}[htp]
    \centering
    \includegraphics[width=\columnwidth]{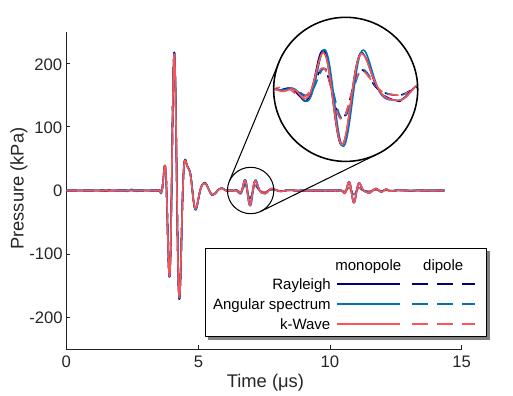}
    \caption{
        \textbf{Comparison of propagation methods and source representations}.
        On-axis simulated pressure at 5~mm distance from a linear array.
    }
    \label{fig:equivalence}
\end{figure}

In k-Wave, or other grid-based collocation methods, the continuous source distributions $S_\mathrm{M}$ and $\mathbf{S}_\mathrm{F}$ are projected onto the grid by convolving them with a band-limited delta function $b$:
\begin{multlinetwocolumn}\label{eq:off-grid-mass}
    \tilde{S}_\mathrm{M}(\mathbf{r},t) = 
    \int_{\mathbb{R}^3}\frac{b(\mathbf{r},\mathbf{r}_1)}{\Delta x \Delta y \Delta z}S_\mathrm{M}(\mathbf{r}_1,t)d^3r_1 = \twocolumnbreak
    2\rho_0\int_{\Sigma_1}\frac{b(\mathbf{r},\mathbf{r}_1)}{\Delta x \Delta y \Delta z}u_n(\mathbf{r},t)d\Sigma_1
\end{multlinetwocolumn}
\begin{multlinetwocolumn}\label{eq:off-grid-force}
    \mathbf{\tilde{S}}_\mathrm{F}(\mathbf{r},t) = \int_{\mathbb{R}^3}\frac{b(\mathbf{r},\mathbf{r}_1)}{\Delta x \Delta y \Delta z}\mathbf{S}_\mathrm{F}(\mathbf{r}_1,t)d^3r_1 = \twocolumnbreak
    \frac{2}{\rho_0}\int_{\Sigma_1}\frac{b(\mathbf{r},\mathbf{r}_1)}{\Delta x \Delta y \Delta z}p(\mathbf{r},t)\mathbf{n_1}d\Sigma_1,
\end{multlinetwocolumn}
with the band-limited delta function as defined in~\cite{Wise2019} and $\Delta x$, $\Delta y$, $\Delta z$ the grid spacing in each direction. 
The numerical source input structures in k-Wave are called pressure sources and velocity sources, which are defined as~\cite{Treeby2016}
\begin{equation}
    \texttt{source.p} = \frac{c_0 \Delta x}{2} \tilde{S}_\mathrm{M}
\end{equation}
and
\begin{equation}
    \texttt{source.ux} = \frac{\Delta x}{2c_0} \tilde{S}_{\mathrm{F},x},
\end{equation}
respectively, and similarly for \texttt{source.uy} and \texttt{source.uz}. In practice, the integrals in~\eqref{eq:off-grid-mass} and~\eqref{eq:off-grid-force} are approximated by a discrete sum. 
If the source lies on the surface $x=0$ and the integration points of the discrete sum coincide with the k-Wave grid, then $\tilde{S}_\mathrm{M}(\mathbf{r},t) = 2\rho_0u_x(\mathbf{r},t)/\Delta x$ and $\tilde{S}_{\mathrm{F},x}(\mathbf{r},t) = 2p(\mathbf{r},t)/(\rho_0\Delta x)$ on the grid points on the source surface and zero elsewhere.
In that case,
\begin{equation}
    \texttt{source.p} = c_0\rho_0u_x
\end{equation}
\begin{equation}
    \texttt{source.ux} = \frac{p}{\rho_0 c_0},
\end{equation}
showing that, somewhat counter-intuitively, the Rayleigh integral over velocity~\eqref{eq:rayleigh-fwd-vp} is equivalent to a pressure source definition in k-Wave and the Rayleigh integral over pressure~\eqref{eq:rayleigh-fwd-pp} to a velocity source definition.

Figure~\ref{fig:equivalence} shows the simulated pressure for both monopole and dipole source definitions for three propagation methods.
The virtual transducer that was used is the default transducer in PROTEUS~\cite{Blanken2024}.
The sensor point was placed on-axis, 5~mm from the transducer surface.
The figure shows that the forward travelling primary wavefront is nearly identical for monopole and dipole sources.
Two edge waves are also visible: the first edge wave originates from the transducer edges along the shorter dimension of the transducer (elevation direction), while the second edge wave originates from the edges along the larger dimension (lateral direction).
Contrary to the forward travelling primary wave, the edge waves differ substantially between monopole and dipole source representations, in accordance with the angle dependency in a dipole field.
The figure also shows that the differences between the three propagation methods are small: the maximum difference between the monopole curves is 2.1\% of the maximum amplitude, and the maximum difference between the dipole curves is 1.5\% of the maximum amplitude.

\subsection{Measurement plane orientation}\label{sec:orientation}

In practice, it can be difficult to orient the transducer surface parallel to the hydrophone scan plane.
However, a misalignment of only 1$^\circ$ can already be detrimental to some of the methods that will be introduced next.
For example, if a linear array with a width of 30~mm that is rotated by $1^\circ$ about the elevation axis receives a backward propagated plane wave with a wavelength of 0.5~mm, it will experience a phase difference of $2\pi$ radians between the first and last elements of the array.
Such a phase difference causes operations such as averaging the incoming pressure over the transducer surface to be inaccurate.
However, if we know the orientation of the scan plane, we can feed the rotated measurement points into the backward Rayleigh integrals~\eqref{eq:rayleigh-bwd-pp} and~\eqref{eq:rayleigh-bwd-pv} to compensate fully for the imperfect orientation.
Here, we introduce a method to determine the orientation of the scan plane with respect to the transducer surface using the angular spectrum of the measurement data at a particular frequency component with wave number $k$.

\begin{figure}[htp]
    \centering
    \includegraphics[width=\columnwidth]{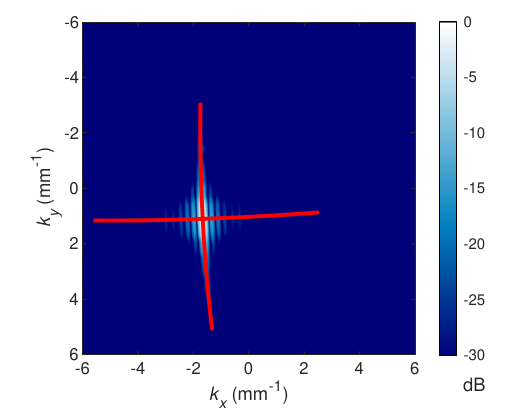}
    \caption{
        \textbf{Estimation of measurement plane orientation.}
        Angular spectrum of the simulated field of a rectangular aperture transducer.
        Ground truth sensor plane orientation: $\theta_x = 6^\circ$, $\theta_y = 9^\circ$, $\theta_z = 3^\circ$.
        The red curves are the best fitting set of integration points.
        Estimated orientation: $\theta_x = 6.04^\circ$, $\theta_y = 9.01^\circ$, $\theta_z = 2.99^\circ$.
    }
    \label{fig:angle-demo}
\end{figure}

A point $(k_x,k_y)$ in the angular spectrum within the region $k_x^2 + k_y^2 \leq k^2$ can be interpreted as the projection of a wave vector $\mathbf{k}$ in $\mathbb{R}^3$ onto the two-dimensional measurement plane.
Let $\mathcal{S}$ be the coordinate system of the unrotated measurement plane and $\mathcal{S}'$ the coordinate system obtained through a rotation represented by an unknown rotation matrix $\mathbf{R}$.
If $\mathbf{k}$ is represented in $\mathcal{S}$ as column vector $[k_x ~ k_y ~ k_z]^\mathrm{T}$ and in $\mathcal{S'}$ as $[k_x' ~ k_y' ~ k_z']^\mathrm{T}$, where $\mathrm{T}$ denotes the transpose, then $[k_x' ~ k_y' ~ k_z']^\mathrm{T} = \mathbf{R}^\mathrm{T}[k_x ~ k_y ~ k_z]^\mathrm{T}$, or
\begin{align}
    k_x' & = R_{11}k_x + R_{21}k_y + R_{31}k_z \\
    k_y' & = R_{12}k_x + R_{22}k_y + R_{32}k_z,
\end{align}
where $R_{ij}$ denote the matrix elements of $\mathbf{R}$.

Transducer characterization will typically be conducted with plane-wave, zero-angle transmissions.
In a perfectly aligned measurement plane, the strongest component in the angular spectrum will therefore correspond to ($k_{x,\mathrm{max}}$, $k_{y,\mathrm{max}}$) = (0, 0), or to the wave vector $[0 ~ 0 ~ k]^\mathrm{T}$ in $\mathcal{S}$.
In a rotated measurement plane, this strongest component will appear at ($k_{x,\mathrm{max}}'$, $k_{y,\mathrm{max}}'$) = ($kR_{31}$, $kR_{32}$).

To proceed, it is useful to decompose the rotation $\mathbf{R}$ into
a rotation of $\theta_x$ about the $x$ axis, followed by
a rotation of $\theta_y$ about the $y$ axis, followed by
a rotation of $\theta_z$ about the $z$ axis, i.e. $\mathbf{R} = \mathbf{R}_z\mathbf{R}_y\mathbf{R}_x$.
Then,
\begin{align}
    k_{x,\mathrm{max}}' & = -k\sin(\theta_y)\\
    k_{y,\mathrm{max}}' & = k\sin(\theta_x)\cos(\theta_y),
\end{align}
which can be solved to find $\theta_x$ and $\theta_y$.



To find the remaining angle $\theta_z$, i.e., the rotation about the propagation axis of the transducer, we can look at salient features in the angular spectrum.
Here, we will consider a transducer with a rectangular aperture.
Such apertures will have an angular spectrum with bright components along the lines $k_x=0$ and $k_y=0$.
These lines can be represented as a parametric set of wave vectors, expressed in $\mathcal{S}$ as
\begin{align}
    \left\{
    [k\sin{\alpha},0,k\cos{\alpha}]^\mathrm{T} \mid
    \alpha\in\mathbb{R}
    \right\} \\
    \left\{
    [0,k\sin{\beta},k\cos{\beta}]^\mathrm{T} \mid
    \beta\in\mathbb{R}
    \right\}
\end{align}
In $\mathcal{S'}$, this set is expressed as
\begin{multlinetwocolumn}\label{eq:rotated-points-x}
    \left\{
    [q_x^{(1)}(\alpha),q_y^{(1)}(\alpha),q_z^{(1)}(\alpha)]^\mathrm{T} \equiv
    \right. \twocolumnbreak \left.
    \mathbf{R}^\mathrm{T}([k\sin{\alpha},0,k\cos{\alpha}]^\mathrm{T} \mid
    \alpha\in\mathbb{R}
    \right\}
\end{multlinetwocolumn}
\begin{multlinetwocolumn}
    \left\{
    [q_x^{(2)}(\beta),q_y^{(2)}(\beta),q_z^{(2)}(\beta)]^\mathrm{T} \equiv
    \right. \twocolumnbreak \left.
    \mathbf{R}^\mathrm{T}[0,k\sin{\beta},k\cos{\beta}]^\mathrm{T} \mid
    \beta\in\mathbb{R}
    \right\}\label{eq:rotated-points-y}
\end{multlinetwocolumn}

Since we have already solved for $\theta_x$, and $\theta_y$, we can write $\mathbf{R} = \mathbf{R}(\theta_z)$.
Now, we can find $\theta_z$ by solving the optimization problem
\begin{multline}\label{eq:rotation-optimization}
    \theta_z = \arg\max_{\theta_z'}
    \left[
    \int_{-\alpha_1}^{\alpha_1} \left|\hat{P}'
        \left(
            q_x^{(1)}(\alpha;\theta_z'),
            q_y^{(1)}(\alpha;\theta_z')
        \right)\right|^2d\alpha
    \right.
    \\ +
    \left.
    \int_{-\beta_1}^{\beta_1} \left|\hat{P}'
        \left(
            q_x^{(2)}(\beta;\theta_z'),
            q_y^{(2)}(\beta;\theta_z')
        \right)\right|^2d\beta
    \right].
\end{multline}
Suitable integration limits $\alpha_1$ and $\beta_1$ can be chosen based on the signal-to-noise ratio of the angular spectrum $\hat{P}'(k_x',k_y')$.

Figure~\ref{fig:angle-demo} illustrates the method.
Rayleigh integral~\eqref{eq:rayleigh-fwd-vp} was used to propagate the source of Fig.~\ref{fig:angular-spectrum}a to a plane at 5~cm from the source rotated by $\theta_x = 6^\circ$, $\theta_y = 9^\circ$, and $\theta_z = 3^\circ$, respectively.
The red curves represent the set of integration points from~\eqref{eq:rotated-points-x} and~\eqref{eq:rotated-points-y} that maximize~\eqref{eq:rotation-optimization}.
The integral of the angular spectrum over these integration points is maximized for $\theta_z = 2.99^\circ$,
with $\theta_x = 6.04^\circ$ and $\theta_y = 9.01^\circ$.

\subsection{Data acquisition}\label{sec:data-acquisition}

Data acquisition comprises two parts: i) measurement of the acoustic output field with a hydrophone and ii) measurement of the element receive voltage in response to the field reflected by a metal plate.
The P4-1 transducer (Philips ATL) was connected to a Vantage 256 system (Verasonics, Kirkland, WA, USA). To approximate the transmit impulse response of the transducer, it was driven with a 12-ns, 30-V rectangular pulse.
Uniform apodization was used for the full transducer aperture, with no electronic delays.
The transducer was interfaced with a tank filled with degassed water using acoustically transparent polymer film and ultrasonic gel.
A fibre-optic needle hydrophone with a sensor diameter of 10~µm (Precision Acoustics, Dorchester, United Kingdom) was mounted onto an xyz linear motorized stage (Optics Focus Instruments Co., Ltd., Beijing, China).
The output of the fibre-optic hydrophone system was connected to a PicoScope 5000 A digital oscilloscope (Pico Technology, St Neots, United Kingdom) with a 15-MHz low-pass filter and a 50-Ω RF terminator. The sampling rate of the oscilloscope was set to 25 MS/s.

As fibre-optic needle hydrophones have a high noise-equivalent pressure~\cite{Hurrell2012,Morris2009a}, the acoustic pressure was averaged over a 1000-pulse repetition sequence for each measurement position to achieve an adequate signal-to-noise ratio.
At each measurement position, the transmit sequence was triggered with a 577 Pulse Generator (Berkeley Nucleonics, San Rafael, CA, USA).
In turn, the Vantage 256 sent a trigger pulse to the oscilloscope synchronized with each transmit pulse.
The oscilloscope, stage controller, and pulse generator were controlled with a custom MATLAB program.

The step size of the motorized stage was set to 0.15~mm to ensure Nyquist sampling of the highest frequency component of the transducer output (about 5~MHz, 0.3~mm wavelength).
The tip of the needle hydrophone was placed at about 3~mm from the transducer surface.
This is close enough to the transducer to neglect the effects of nonlinear propagation and far enough to ensure that evanescent waves are sufficiently attenuated.
Evanescent waves can provide sub-wavelength resolution information about the source~\cite{Lee1985} but require an even smaller step size to prevent aliasing.
As the frontal area of the transducer is 20 by 30~mm, the scan plane size was set to 24 by 36~mm to ensure that the full pressure field would be captured by the hydrophone.

To obtain the receive characteristics of the transducer, a metal plate was placed in front of the transducer surface to project the transmitted field back onto the transducer surface.
We determined the pressure reflection coefficient of the plate to be $R_p=0.78$.
The element receive data of the Vantage 256 system was recorded with the time-gain compensation set to zero.

\begin{figure*}[htp]
    \centering
    \includegraphics[width=13cm]{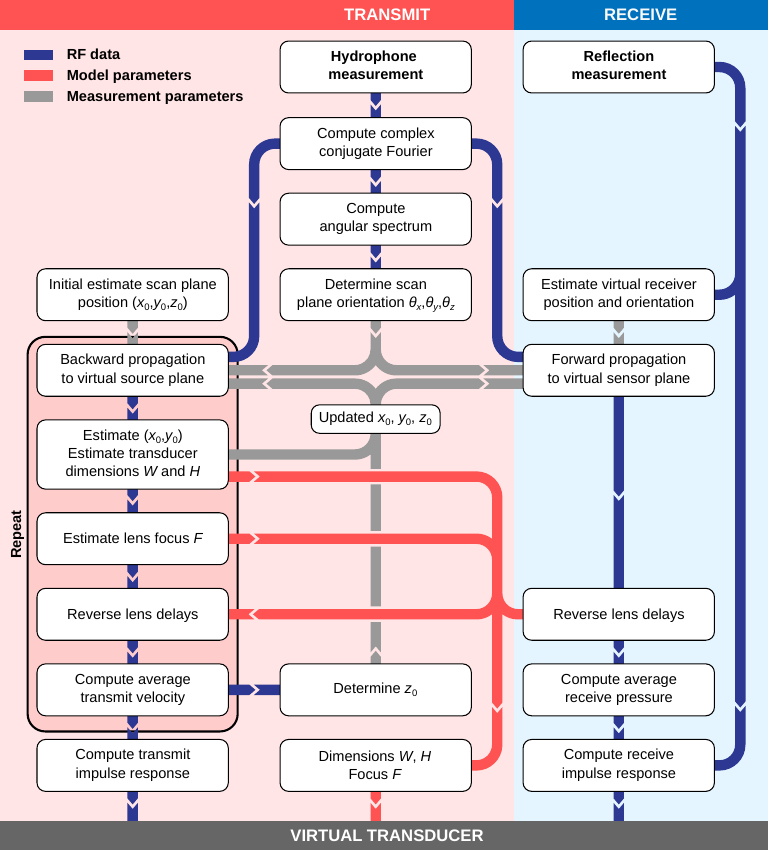}
    \caption{
        \textbf{Data processing pipeline.}
        From data acquisition to virtual transducer definition.
    }
    \label{fig:pipeline}
\end{figure*}

\subsection{Data processing pipeline}\label{sec:pipeline}

Fig.~\ref{fig:pipeline} outlines the data processing pipeline from the acquired radio-frequency (RF) data to the construction of a virtual transducer.
In this section, we will briefly describe each of the processing steps.
In Section~\ref{sec:results}, we will demonstrate the application of these steps to the acquired RF data and provide further details.

First, the complex conjugate Fourier transform of the hydrophone data with respect to time is computed~\eqref{eq:complex-conjugate-fourier} as many of the subsequent operations are defined in the frequency domain.
The hydrophone voltage RF data are converted to pressure RF data using the sensitivity curve of the calibrated hydrophone.
After computing the angular spectrum of the measurement plane~\eqref{eq:angular-spectrum}, the orientation of the measurement plane is determined with the method described in Section~\ref{sec:orientation}.

With the estimate of the measurement plane orientation and an initial guess for the measurement plane centre ($x_0$, $y_0$, $z_0$), the pressure in the measurement plane is propagated backward to the surface $z=0$.
In our demonstration, we choose to use a Rayleigh integral~\eqref{eq:rayleigh-bwd-pv} for the backward propagation, for the reasons described in Section~\ref{sec:rayleigh-integrals}.
The Rayleigh integral is evaluated as a sum over the measurement points.
As shown in~\cite{Sapozhnikov2015}, taking the integration step size to be equal to the Nyquist sampled measurement step size, does not introduce additional error, provided the planes are sufficiently separated from each other.
The width $W$ and height $H$ of the transducer aperture are determined by fitting a rectangular aperture to the amplitude of the reconstructed source.
Based on the centre of the fitted aperture ($x_\mathrm{c}$, $y_\mathrm{c}$), the values of $x_0$ and $y_0$ can be updated: $x_0 \leftarrow x_0 - x_\mathrm{c}$ and $y_0 \leftarrow y_0 - y_\mathrm{c}$.
The focal distance $F$ of the acoustic lens can be extracted from the phase delays of the reconstructed source.
The acoustic lens effectively applies delays $\tau$ along the length of the transducer elements $y$:
\begin{equation}\label{eq:lens-delays}
    \tau = (\sqrt{(H/2)^2+F^2} - \sqrt{y^2+F^2})/c_0.
\end{equation}
After determining $F$, the lens delays are reversed, temporally aligning the source signals across the transducer surface and allowing the computation of the average transmit velocity.
The onset time $t_0$ of this signal can be used to update the estimated distance between source plane and measurement plane: $z_0 \leftarrow z_0 + c_0t_0$.
With the updated values of $x_0$, $y_0$, and $z_0$, the steps in the shaded box in Fig.~\ref{fig:pipeline} are repeated.
Finally, the transmit impulse response is computed by deconvolving the average transmit velocity with the driving signal.
Since we have used a short delta-like pulse for the driving signal, \eqref{eq:transmit-impulse-response} can be approximated by
\begin{equation}\label{eq:deconvolution-transmit}
    u_\mathrm{T}(t) = h_\mathrm{T}(t)\int E_\mathrm{T}(t') dt'.
\end{equation}
Therefore, the deconvolution simplifies to a division by the area under the curve of the driving signal.

To determine the receive impulse response~\eqref{eq:receive-impulse-response}, the pressure that is reflected back to the transducer needs to be computed.
Computing the reflected pressure is equivalent to multiplying the pressure field by the reflection coefficient and forward propagating the field to the mirror image of the transducer, which we will refer to as the virtual sensor.
The orientation and position of the virtual sensor can be estimated from the element receive data.
Similar to the virtual source data, the lens delays are reversed for the virtual sensor data, and the pressure is averaged over the sensor surface to obtain the average receive pressure $p_\mathrm{R}(t)$.
The receive impulse response is then obtained through deconvolution in the frequency domain.


\section{Results}\label{sec:results}

\begin{figure}[htp]
    \centering
    \includegraphics[width=\columnwidth]{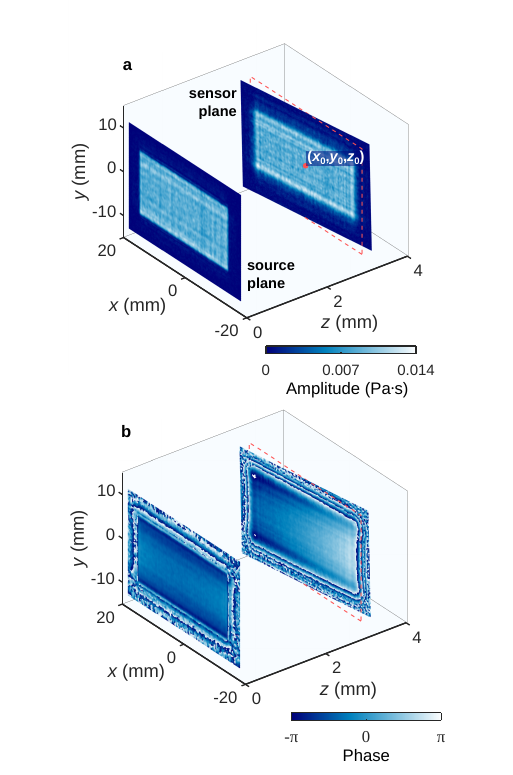}
    \caption{
        \textbf{Backward propagation from the sensor plane to the source plane.}
        First iteration of backward propagation.
        \textbf{a} Amplitude of the 2.5-MHz component of the measurement data and the reconstructed source.
        The red dot is the estimated centre of the measurement plane. 
        The red dashed line represents the measurement plane before applying the rotation.
        \textbf{b} Phase of the 2.5-MHz component.
    }
    \label{fig:backpropagation}
\end{figure}

\subsection{Transmit characteristics}\label{sec:transmit}

In this section we will apply the data processing pipeline from Section~\ref{sec:pipeline} to the data obtained in Section~\ref{sec:data-acquisition}.
Using the method described in Section~\ref{sec:orientation}, we find that the measurement plane had an orientation that can be described by the rotations $\theta_x=-0.08^\circ$, $\theta_y=0.65^\circ$, and $\theta_z=-0.77^\circ$, respectively.
In practice, the transducer was aligned with respect to a fixed measurement system.
However, since we define the coordinate system with respect to the transducer, we are writing about the orientation of the measurement plane instead.
Based on visual inspection of the setup, the initial guess for the centre of the measurement plane is $(x_0,y_0,z_0)$ = (0, 0, 3.0~mm).
Figure~\ref{fig:backpropagation}a shows the orientation of the measurement plane for this initial estimate.
The pressure data is propagated backward to the transducer surface with the Rayleigh integral in~\eqref{eq:rayleigh-bwd-pv}.
We use $\rho_0 = 998$~kg/m$^3$ and $c_0 = 1481$~m/s (the values for water at $20^\circ$C).
The result for the 2.5~MHz component is shown in Fig.~\ref{fig:backpropagation}.
For visualization purposes, the reconstructed source velocity $V$ is multiplied by $\rho_0c_0$ to match the units of the measurement data (Pa$\cdot$s in the frequency domain).
The backward propagation is only applied to the positive frequency components $P(\mathbf{r};f)$ to limit the computational cost.
The negative frequency components $P(\mathbf{r};-f)$ can simply be reconstructed with $P(\mathbf{r};-f) = P(\mathbf{r};f)^*$, based on the fact that the time-domain signals must be real.
Figure~\ref{fig:backpropagation}b shows the corresponding phase of the 2.5~MHz component.
The measurement plane shows a strong linear gradient in the phase, which is due to the inclination of the measurement plane relative to the wavefront.
This lateral gradient is absent in the source plane, demonstrating an effective compensation of the misalignment.
Only a parabolic gradient remains, which is due to the acoustic lens.

\begin{figure*}[htp]
    \centering
    \includegraphics[width=13cm]{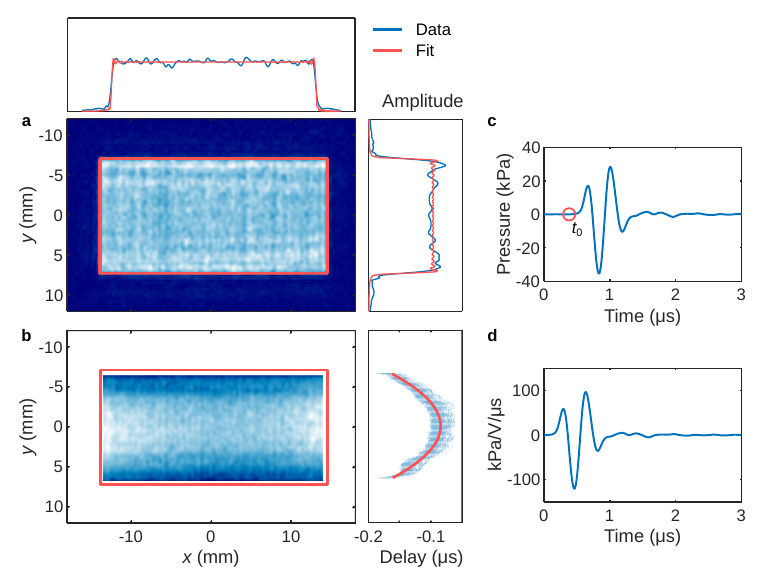}
    \caption{
        \textbf{Model parameter and transmit impulse response estimation}
        \textbf{a} Best fit to the amplitude of the 2.5-MHz component of the reconstructed source plane (first iteration).
        \textbf{b} Best fit to the phase delay of the 2.5-MHz component (first iteration).
        \textbf{c} Pressure averaged over the source surface after reversing the lens delays.
        The red circle indicates the estimated start time $t_0$ of the transmit pulse.
        \textbf{d} Estimated impulse response of the transducer after the second iteration of backward propagation.
    }
    \label{fig:model-parameters}
\end{figure*}

To find the width $W$ and the lateral centre $x_\mathrm{c}$, we fit a band-limited rectangle function to the data, because the experimental data is also band-limited (Section~\ref{sec:angular-spectrum}).
A band-limited rectangle function can be obtained by band-limiting a sinc function in $k$-space and taking the inverse Fourier transform:
\begin{equation}
    V_\mathrm{fit}(x) = \frac{V_0W}{2\pi}\int_{-k}^{k}\mathrm{sinc}\left(\frac{k_xW}{2\pi}\right)e^{ik_x(x-x_\mathrm{c})}dk_x,
\end{equation}
with $\mathrm{sinc}(\alpha) = \sin(\pi\alpha)/(\pi\alpha)$.
Here, $V_0$, $W$, and $x_\mathrm{c}$ are fitting parameters.
We apply a similar fit to find the height $H$ and the elevation centre $y_\mathrm{c}$.
The least-square fits are shown in Fig.~\ref{fig:model-parameters}a.

To find the focal distance $F$ of the acoustic lens, we apply a parabolic fit to the phase delays $\tau$ (phase multiplied by angular frequency $\omega$).
In the paraxial approximation, \eqref{eq:lens-delays} can be approximated by
\begin{equation}\label{eq:lens-delays-paraxial}
    \tau = \frac{(H/2)^2}{2Fc_0} - \frac{y^2}{2Fc_0}.
\end{equation}
Figure~\ref{fig:model-parameters}b shows the least-squares parabolic fit.
The focal distance $F$ can be computed from the curvature of the parabola.
Careful inspection of Fig.~\ref{fig:model-parameters}b does not only show curvature along the $y$-axis, but also curvature along the $x$-axis.
This second curvature may be due to imperfections during the fabrication of the less.
As it is much weaker, we neglect it in our model.

Next, the lens delays are reversed by multiplying the frequency domain data by $\exp(i\omega\tau)$, the frequency domain data is transformed back to the time domain, and the average is taken over the aperture of the source plane.
The resulting waveform is shown in Fig.~\ref{fig:model-parameters}c (multiplied by $\rho_0c_0)$.
From this curve, we can extract the onset time $t_0$ of the signal.
We approximate the onset time as the first zero-crossing before the signal exceeds the noise level.
Based on $t_0$, $x_c$, and $y_c$, we find the updated centre of the measurement plane $x_0,y_0,z_0$ = ($-0.31$~mm, $-0.06$~mm, $3.57$~mm).
With these updated values, we repeat the operations in the shaded box in Fig.~\ref{fig:pipeline}.
In this second iteration, we find $W = 28.54$~mm, $H=14.28$~mm, and $F = 181$~mm.
Finally, we divide the average transmit velocity by the area under the curve of the transmit voltage~\eqref{eq:deconvolution-transmit}, which is 0.36~µs$\cdot$V.
The resulting transmit impulse response (multiplied by $\rho_0c_0$) is shown in Fig.~\ref{fig:model-parameters}d.

\subsection{Receive characteristics}\label{sec:receive}

\begin{figure*}[htp]
    \centering
    \includegraphics[width=\textwidth]{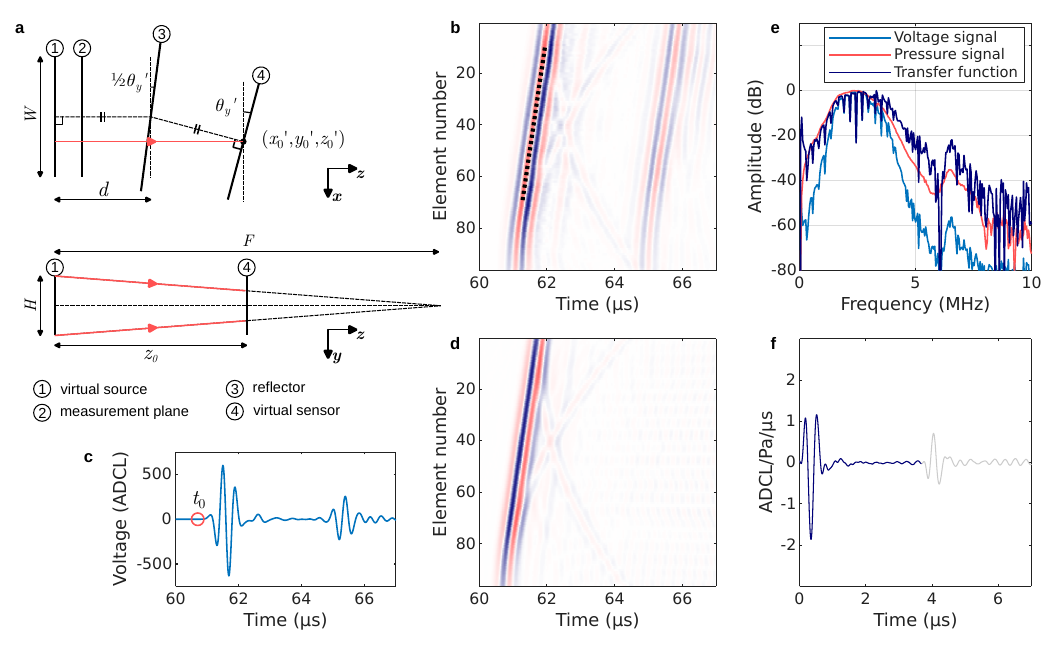}
    \caption{
        \textbf{Receive transfer function estimation.} 
        \textbf{a} Position and orientation of the virtual source, the measurement plane, the reflector, and the virtual sensor.
        The red line shows the fastest path between source and sensor.
        \textbf{b} Measured transducer element receive voltage data.
        \textbf{c} Averaged receive data.
        \textbf{d} Transducer element receive pressure data obtained through forward propagation.
        \textbf{e} Deconvolution of the receive voltage signal in the frequency domain.
        \textbf{f} Estimated receive impulse response.
    }
    \label{fig:receive}
\end{figure*}

Figure~\ref{fig:receive}a shows a schematic of the receive setup.
The front surface of the metal plate is placed at a distance $d$ from the virtual transducer surface and rotated by an unknown and unintended angle $\frac{1}{2}\theta_y'$ about the $y$-axis.
The transmitted wave is reflected back to the virtual source at an angle $\theta_y'$.
As the metal plate acts as a specular reflector, this situation is equivalent to a forward propagation from the virtual source to its mirror image, which we refer to as the virtual sensor, provided that we multiply the field amplitude by the reflection coefficient $R_p=0.78$ and the angles of the incident $k$-vectors are small enough to consider the reflection coefficient to be constant.

The point $(x_0',y_0',z_0')$ represents the centre of this virtual sensor.
The angle $\theta_y$ can be determined from the element receive voltage RF data in Fig.~\ref{fig:receive}b.
The dashed line is a linear fit to the peaks of the main wavefront.
The slope of this fit is $-p_\mathrm{el}\tan\theta_y'/c_0$, with $p_\mathrm{el}$ the element pitch, yielding an angle $\theta_y = 3.17^\circ$.
Figure~\ref{fig:receive} also shows a second, parallel wavefront starting around $t=65$~µs, which is the reflection from the back surface of the metal plate.
In principle, the virtual sensor may also be rotated by an angle $\theta_x'$ about the $x$-axis, but this angle cannot be determined from the receive data.
For lack of a better method, we will therefore assume that the reflector was perfectly aligned about this axis and $\theta_x'=0$.

To find $z_0'$, we apply a similar method as in Section~\ref{sec:transmit}.
First, we reverse the delays in Fig.~\ref{fig:receive}b, such that the peaks of all RF signals align, and compute the average over all RF signals, shown in Fig.~\ref{fig:receive}c.
The voltage is expressed in analogue-to-digital converter levels (ADCL).
As before, we estimate the onset time $t_0$ as the first zero-crossing before the pulse exceeds the noise level.
The onset time comprises two components: i) the time corresponding to the shortest travel path of the wavefront between the virtual source and the virtual sensor, indicated by the red lines in Fig.~\ref{fig:receive}a and ii) an additional delay due to the lens on receive~\eqref{eq:lens-delays}.
From~\eqref{eq:lens-delays} and the geometry in Fig.~\ref{fig:receive}a, it follows that
\begin{multlinetwocolumn}
    t_0 = \frac{1}{c_0}\left[
    \left(1 + \frac{z_0'}{F}\right)\sqrt{(H/2)^2+F^2}
        \right. \twocolumnbreak \left.
    - \sqrt{(1-z_0'/F)^2(H/2)^2 + F^2}
    \right].
\end{multlinetwocolumn}
Solving for $z_0'$, gives $z_0'=89.73$~mm.
From Fig.~\ref{fig:receive}a it follows that $d = z_0'/(1+\cos\theta_y') = 44.90$~mm and $x_0'= d\sin\theta_y' = 2.48$~mm.

The next step is forward propagating the pressure field from the measurement plane to the virtual sensor.
The orientation of the measurement plane, determined in Section~\ref{sec:transmit}, must be taken into consideration again.
In principle, the pressure field could also be propagated from the virtual source to the virtual sensor.
However, the virtual source was obtained through backward propagation from the measurement plane in Section~\ref{sec:transmit}, meaning that the Rayleigh integral would need to be applied a second time, leading to propagation of numerical error.
After forward propagation, the lens delays~\eqref{eq:lens-delays} are reversed and the pressure is averaged over the length $H$ of the transducer elements.
The resulting pressure RF data is shown in Fig.~\ref{fig:receive}d.
The second wavefront around $t = 65$~µs is not visible because the reflection from the back surface of the reflector was not modelled.
Analogously to the voltage receive data, we correct for the inclination and compute the average of all transducer elements.

With both the voltage and pressure receive data at hand, the receive impulse response~\eqref{eq:receive-impulse-response} can be determined.
We obtain the receive impulse response through deconvolution in the frequency domain.
The frequency domain equivalent of~\eqref{eq:receive-impulse-response} is
\begin{equation}
    Y_\mathrm{R}(f) = H_\mathrm{R}(f)P_\mathrm{R}(f),
\end{equation}
with $Y_\mathrm{R}(f)$, $H_\mathrm{R}(f)$, and $P_\mathrm{R}(f)$ the Fourier transforms of $E_\mathrm{R}(t)$, $h_\mathrm{R}(t)$, and $p_\mathrm{R}(t)$, respectively.
Computing the inverse of this equation is an ill-posed problem because the signals $Y_\mathrm{R}(f)$ and $P_\mathrm{R}(f)$ are both corrupted by noise.
For a noisy system defined by $Y_\mathrm{R}(f) = H_\mathrm{R}(f)P_\mathrm{R}(f) + N_Y(f)$, $H_\mathrm{R}$ can be approximated using Wiener deconvolution~\cite{Kino1987}:
\begin{equation}
    H_\mathrm{R}(f) = \frac{Y_\mathrm{R}(f)}{P_\mathrm{R}(f)}\frac{1}{1 + \mathrm{SNR}_Y(f)},
\end{equation}
where the signal-to-noise ratio is defined as $\mathrm{SNR}_Y = |Y_\mathrm{R}|^2/|N_Y|^2$.
However, in the current system, both $Y_\mathrm{R}(f)$ and $P_\mathrm{R}(f)$ are noisy.
Therefore, we modify the Wiener deconvolution as follows:
\begin{equation}\label{eq:deconvolution-receive}
    H_\mathrm{R}(f) = \frac{Y_\mathrm{R}(f)}{P_\mathrm{R}(f)}\frac{1}{1 + \mathrm{SNR}_Y(f) + \mathrm{SNR}_P(f)},
\end{equation}
where $\mathrm{SNR}_P = |P_\mathrm{R}|^2/|N_P|^2$ acts as a regularization term on the noise in $P_\mathrm{R}(f)$.
Figure~\ref{fig:receive}e shows $Y_\mathrm{R}(f)$, $P_\mathrm{R}(f)$, and the computed $H_\mathrm{R}(f)$.
Taking the inverse Fourier transform of $H_\mathrm{R}$ yields the receive impulse response, which is shown in Fig.~\ref{fig:receive}f.
Finally, the signal is truncated (blue curve) to eliminate the second and subsequent pulses in the signal that are the result of the secondary reflections (grey signal).

\subsection{Model validation}\label{sec:validation}

\begin{figure}[htp]
    \centering
    \includegraphics[width=\columnwidth]{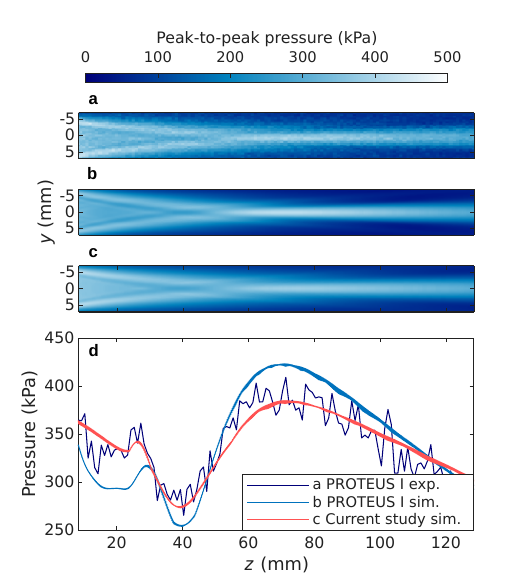}
    \caption{
    \textbf{Validation of the virtual transducer on independent experimental data.}
        \textbf{a} Measured field of the transducer from~\cite{Blanken2024}.
        \textbf{b} PROTEUS simulation of the field from~\cite{Blanken2024}.
        \textbf{c} PROTEUS simulation of the field with the virtual transducer obtained in the current work.
        \textbf{d} On-axis pressure profiles from the fields in \textbf{a}-\textbf{c}.
        The rapid, small-amplitude oscillations ($\sim 5$~kPa) in the simulated data are due to sampling effects.
    }
    \label{fig:comparison}
\end{figure}

We validate the virtual transducer model on experimental data that was not used to construct the model.
To that end, we simulate the experimental pressure field presented in PROTEUS Part~I~\cite{Blanken2024}, which is displayed in Fig.~\ref{fig:comparison}a.
In this experiment, a 7.5-V, 2.5-MHz, 2-cycle driving signal was used, a cosine-tapered apodization window was applied, and electronic delays were applied to create a virtual focus point at -35~mm from the transducer.
In PROTEUS Part~I, the field in Fig.~\ref{fig:comparison}a was used to find the focal distance of the lens and the absolute amplitude of the impulse response of the P4-1 transducer by matching simulation and experiment.
The simulated field from PROTEUS Part~I has been reproduced in Fig.~\ref{fig:comparison}b.
Figure~\ref{fig:comparison}c was created by importing the virtual transducer from the current study into the PROTEUS simulator and simulating the field with the same transmit settings.
In Fig.~\ref{fig:comparison}d, the on-axis pressure profiles of all three fields are displayed.

The pressure field simulated with the new virtual transducer was multiplied by a factor of 1.2 to obtain the best fit to the data.
We provide two explanations why the new simulation needs to be rescaled. 
i) The experimental data in Fig.~\ref{fig:comparison}a was measured with a different fibre-optic needle hydrophone, and there exists roughly a 10\% uncertainty in the sensitivity data of these hydrophones.
ii) The virtual transducer in the current study was characterized with a 12-ns rectangular driving pulse.
This is the minimum pulse duration that the Vantage 256 system can provide.
Below this value, the output becomes unreliable.
Therefore, the driving pulse might not have been perfectly rectangular, and we might have overestimated the area under the curve in~\eqref{eq:deconvolution-transmit}.

Despite the slight mismatch in absolute amplitude, it is clear that the current characterization provides a more accurate fit to the \emph{shape} of the pressure profile than the characterization in PROTEUS Part~I.
In that study, the elevation focus was estimated to be 11~cm, whereas, in the current study, we found an elevation focus of 18~cm.
This discrepancy is related to the height $H$ of the transducer aperture.
In the current study, we determined the height to be $H=14.28$~mm.
However, in PROTEUS Part~I, $H=16$~mm was used, which is the value provided by the manufacturer.
This overestimation of the transducer height resulted in a larger natural focal distance, which has the effect of moving the peak intensity point further away from the transducer.
In turn, the focal distance of the lens was underestimated to keep the peak intensity point at the same position as in the experimental data.

\subsection{Spurious modes of oscillation}

\begin{figure*}[htp]
    \centering
    \includegraphics[width=\textwidth]{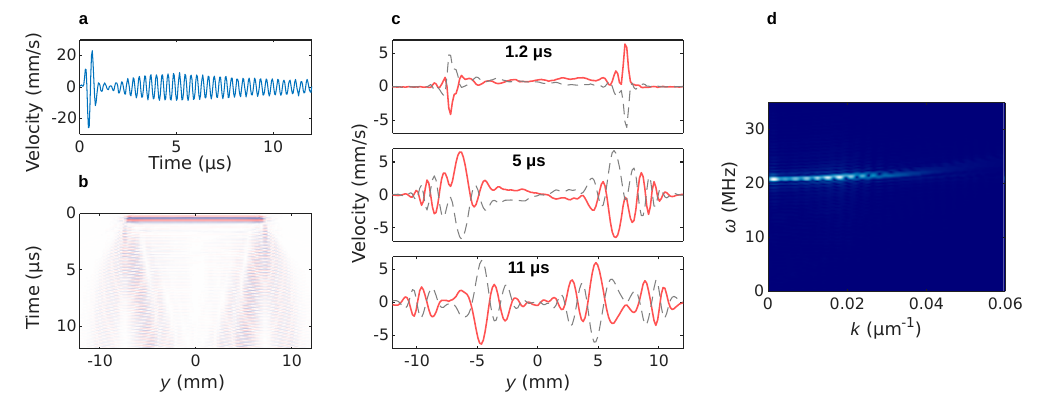}
    \caption{
        \textbf{Spurious mode of oscillation.}
        \textbf{a} Normal velocity of the transducer surface close to the edge of the transducer.
        \textbf{b} Normal velocity along the length the transducer elements. Averaged over all transducer elements.
        \textbf{c} Normal velocity at selected time points from \textbf{b}.
        \textbf{d} Dispersion plot of the spurious mode.
    }
    \label{fig:modes}
\end{figure*}

In the formulation of the transmit impulse response~\eqref{eq:transmit-impulse-response}, we assume that the piezo-electric elements of the transducer oscillate in thickness expander mode~\cite{Krimholtz1972,Stepanishen1981}, meaning that the surface of the elements oscillates uniformly in the direction perpendicular to the transducer surface.
However, many other modes of oscillation are known to exist~\cite{Larson1981}.
In the reconstructed source obtained in Section~\ref{sec:transmit}, we also find evidence for another mode of oscillation.
Figure~\ref{fig:modes}a shows the reconstructed normal velocity at the transducer surface close to an edge of the acoustic aperture.
The broadband 2.5-MHz excitation pulse is followed by a narrowband 3.3~MHz tail.
This tail has been described before in~\cite{Blanken2024}, where it was assumed to be a measurement artifact of the hydrophone system.

Here, we demonstrate that this signal cannot be a measurement artifact but must be a spurious mode of oscillation.
Figure~\ref{fig:modes}b visualizes the surface velocity as a function of time, averaged across the $x$-axis.
In Fig.~\ref{fig:modes}c, cross-sections of this data at selected time points are shown, clearly showing an antisymmetric oscillation mode.
As a fibre-optic needle hydrophone has circular symmetry around the propagation axis~\cite{Hurrell2012,Morris2009a}, the 3.3~MHz signal cannot be discarded as a measurement artifact.
Interestingly, the antisymmetric mode exists primarily along the length of the transducer elements ($y$-direction).
Averaging across the $y$-axis still shows some antisymmetric oscillations, but the effect is much weaker.
This may explain why the 3.3~MHz tail is not visible in the element receive data in Fig.~\ref{fig:receive}b, as in receive, antisymmetric oscillations average out to zero.

Another interesting observation is that the waves in Fig.~\ref{fig:modes}b do not only move inward, but also outward, beyond the boundaries of the active aperture.
This may suggest that the piezoelectric elements extend beyond the electrodes or that the spurious mode is supported by a different layer of the transducer.

In addition to the spurious waves at 3.3~MHz, Fig.~\ref{fig:model-parameters}a reveals spatial fluctuations in the amplitude of the 2.5-MHz component of the signal.
These fluctuations may also indicate a deviation from the uniform oscillations predicted by the thickness-expander model.
However, they may also be due to changes in the driving voltage during the hydrophone measurement.
As mentioned in Section~\ref{sec:validation}, the 12-ns driving pulse may have been too short for the Vantage 256 system to supply a reliable transmit voltage.


\section{Discussion}

This work has demonstrated our characterization pipeline on a P4-1 transducer.
We have successfully determined the receive and transmit impulse responses.
The lens focus was found to be $F = 0.181$~m.
The width and height of the transducer aperture were found to be $W = 28.54$~mm and $H=14.28$~mm, respectively.
The width is close to the value of 28.27~mm reported by the manufacturer.
The height, however, is substantially different from the value of 16~mm reported by the manufacturer.
In Section~\ref{sec:validation}, we showed that this mismatch has non-negligible implications for the simulated field.
Furthermore, we have found evidence for a spurious mode of oscillation in the transmit response.
Although we have developed a full characterization pipeline from measurement to virtual transducer, several open questions remain, which we will discuss in this section.

\subsection{Nature of the spurious waves}
We hypothesize that the observed surface waves are Lamb waves (Rayleigh waves propagating in a thin slab of material), based on the following pieces of evidence.
Firstly, the waves originate from the transducer edges.
Delannoy~\textit{et al.} write that ``Lamb waves are launched at the transducer edges, i.e., in the transition zone of the driving electric field, where a gradient of piezoelectric stress exists.''~\cite{Delannoy1980}
Cathignol~\textit{et al.} make a similar statement~\cite{Cathignol1997}.
Secondly, the observed spurious modes are much more pronounced along the transducer elements than across them, in line with the observation by Delannoy~\textit{et al.} that subdividing the source into elements suppresses Lamb waves~\cite{Delannoy1980}.
Thirdly, Lamb waves exist as either symmetric modes or antisymmetric modes~\cite{Delannoy1980,Cathignol1997}.
Finally, Lamb wave modes are dispersive.
Figure~\ref{fig:modes}d shows the dispersion plot of Fig.~\ref{fig:modes}b (both spatial and temporal Fourier transform), showing a nonlinear dispersion curve.
From this curve, it follows that the phase velocity $\omega/k$ is infinite at $k=0$ and rapidly drops with increasing $k$, in line with the dispersion curves presented in~\cite{Delannoy1980} and~\cite{Cathignol1997}.
However, we could not establish a quantitative match between our dispersion curve and dispersion curves in literature as this requires exact knowledge of the thickness of the piezoelectric elements and the type of piezoelectric material.
These spurious modes deserve further attention as it is unclear what their effect on ultrasound imaging quality might be.

\subsection{Reciprocity and the phase response of the transducer}
\begin{figure*}[htp]
    \centering
    \includegraphics[width=\textwidth]{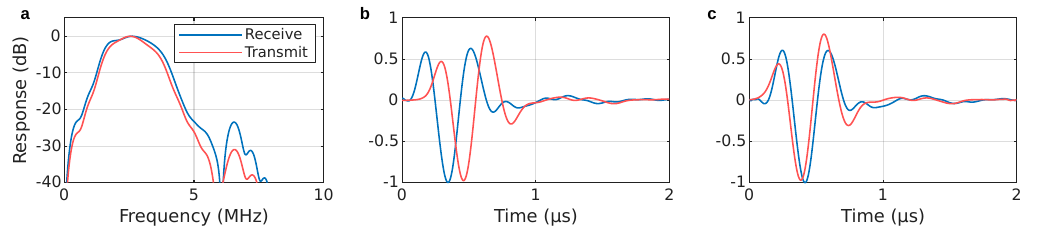}
    \caption{
        \textbf{Phase compensation.}
        \textbf{a} Transmit and receive transfer functions.
        \textbf{b} Normalized transmit and receive impulse responses.
        \textbf{c} Normalized transmit and receive impulse responses after compensating for the phase delays caused by the measurement setup (based on informed assumptions).
    }
    \label{fig:phase}
\end{figure*}

Figure~\ref{fig:phase}a shows the normalized transmit and receive transfer functions (the Fourier transforms of the respective impulse responses).
The curves are surprisingly similar, a result that is reminiscent of the well-known reciprocity theorem for electroacoustic transducers~\cite{Foldy1945}.
However, the normalized transmit and receive impulse responses (Fig.~\ref{fig:phase}b) are not the same.
We identify two sources of phase delay in the system that were note taken into account: i) the analogue 15~MHz low-pass filter and ii) the fibre-optic hydrophone sensor.
We do not have phase calibration data for these components, but we can make an informed estimate of these delays on the determined impulse responses.
For the low-pass filter, we use the phase delay data from a comparable product (BLP-15+, Mini-Circuits, Brooklyn, NY, USA).
For the fibre-optic hydrophone, we assume that it is a minimum-phase system, which allows the phase response to be computed from the amplitude response with Bode's gain-phase relation~\cite{Bechhoefer2011}.
These sources of additional phase delay affect both the transmit impulse response, through~\eqref{eq:deconvolution-transmit}, and the receive impulse response, through~\eqref{eq:deconvolution-receive}, but in opposite directions.
Compensating for these delays results in the impulse responses shown in Fig.~\ref{fig:phase}c.

These results suggest that the normalized time domain responses are also similar.
Nonetheless, we exercise caution before concluding that the normalized impulse responses are exactly the same, though some have assumed so for ease of use~\cite{Meulen2017}.
The reciprocity theorem states that the ratio between the microphone and speaker responses is independent of frequency.
The microphone and speaker responses are related to, but not the same as the transmit and receive transfer functions.
In~\cite{Neer2011}, a relation between the transmit and receive transfer functions is derived for an unfocused circular transducer:
\begin{equation}
    H_\mathrm{R}(\omega)/H_\mathrm{T}(\omega) = 2Z(\omega)A,
\end{equation}
where $Z(\omega)$ is the electrical impedance of the transducer and $A$ is its surface area.
The notion that acoustic transducers can be calibrated using purely electrical measurements has been employed many times~\cite{MacLean1940,Luker1981,Xiao2016}.
Reciprocity calibration could be performed for multi-element medical transducers to verify the transducer responses.
In particular, reciprocity calibration could be used to detect differences between transducer elements.
In the current study, we have assumed the transmit and receive impulse responses to be the same for all elements, but it is known that differences between elements can exist.
Characterizing each of the elements separately with hydrophone measurements would take an impractically long time and would also result in a poor signal to noise ratio.
By contrast, reciprocity calibration allows for rapid characterization of each individual element.
However, acoustic measurements would still be necessary, as reciprocity calibration cannot provide information about the aperture size and lens focus.

\subsection{What is the best source representation?}\label{sec:source-discussion}
In this work, we have modelled the transducer elements as thickness expander elements with a uniform normal velocity.
Implicit in this definition is the assumption that the active aperture is embedded in an infinite rigid baffle (Section~\ref{sec:rayleigh-integrals}).
Although this definition may seem intuitive, there is a priori no compelling reason to adopt it.
We could just as well have used a soft baffle definition.
The choice of source representation primarily affects the edge waves (Fig.~\ref{fig:comparison}).
However, the edge waves in the hydrophone data were too obscured by spurious waves and noise to provide a definitive answer to this question.

Aside from the question which representation is the most physically accurate, there is a practical reason to use the rigid baffle definition in k-Wave (monopole sources).
k-Wave uses a staggered grid~\cite{Treeby2010} for the computation of the acoustic field.
Monopole sources and pressure sensors are placed on the grid points, whereas dipole sources are placed between grid points.
If a dipole transducer source is used, this introduces an offset of half a grid point between the source plane and the sensor plane, which may be a concern for super-resolution applications.
This mismatch could be solved by using an off-grid source definition, but at the cost of a higher memory requirement.

A more pressing question is what happens when the transducer is coupled to a medium other than water.
How do the transmit pressure and velocity change?
The electromechanical coupling depends on many parameters~\cite{Castillo2003}, and this question requires a more in-depth investigation.

\subsection{Orientation of the reflector}
In Section~\ref{sec:orientation}, we have presented a method that is robust to misalignment of the experimental setup.
The full orientation of the measurement plane with respect to the transducer could be extracted from the angular spectrum of the measurement data.
However, for the receive data, only the rotation of the virtual sensor about the $y$-axis could be extracted, because the set of sensor points, i.e. the transducer array, is one-dimensional.
For two-dimensional array, e.g. matrix arrays and row-column arrays, the full orientation of the virtual source ($\theta_x,\theta_y)$ could be extracted from the data (the third angle $\theta_x$ is fixed by mirror symmetry), whereas, for single-element transducers, no information on the orientation could be deduced.
To alleviate this problem, the reflector could be mounted on a tilt stage.
The angles $\theta_x$ and $\theta_y$ could then be fine-tuned while monitoring the receive signal.
The spatially averaged receive signal would be maximized at perfect alignment.

\subsection{Modelling of the acoustic lens}

The virtual transducer surface will, in general, not coincide with the actual surface of the piezoelectric transducer elements because the acoustic lens and the matching layers have a finite thickness.
What matters is that the waves in both simulation and experiment travel the same acoustic path length, which is the case in the paraxial approximation.
However, for large angle transmission and acquisition, the acoustic path difference caused by the lens becomes angle-dependent.
This concern has been pointed out in the context of image reconstruction for spatial compounding~\cite{Choi2009}.
To correct for this lens effect, angle-dependent lens delays could be applied to the virtual source.


\section{Conclusion}

We have developed a data processing pipeline to fully characterize a medical ultrasound transducer and turn it into a virtual transducer that can be used for realistic ultrasound field simulations.
The method is based on a holographic measurement of the pressure field and a measurement of the reflected field in combination with field projections.
The virtual transducer is defined by the transmit impulse response, the receive impulse response, the focal distance of the acoustic lens, and the aperture size.
We have demonstrated the pipeline on a P4-1 phased array transducer.
We found that the height of the transducer does not match the value provided by the manufacturer.
We also found evidence for a spurious oscillation mode that may affect ultrasound imaging quality.
These findings highlight the importance of using holographic measurements for accurate transducer characterization.
The virtual transducer model can be imported into PROTEUS to simulate the ultrasound field and the received RF data for arbitrary transmit settings.
As such, the characterization pipeline serves as an indispensable tool for explorative studies into new ultrasound imaging strategies and for the generation of realistic datasets that can be used in machine-learning-based approaches.
The computational pipeline is available as an open-source toolkit as part of the PROTEUS-SIM software collection.

\begin{acknowledgments}
    The authors acknowledge funding from the 4TU Precision Medicine program
    supported by High Tech for a Sustainable Future,
    a framework commissioned by the four Universities of Technology of the Netherlands.
    Guillaume Lajoinie acknowledges funding from the HORIZON.1.1 Programme of the European Research Council for the
    Super-FALCON project, Grant agreement ID: 101076844
    (DOI: 10.3030/101076844).
\end{acknowledgments}

\appendix*
\section{Averaged spectral propagators}\label{appendix:propagators}

Williams and Maynard have shown that the discrete implementation of the angular spectrum method leads to large bias errors~\cite{Williams1982}. They show that the bias errors can be strongly reduced if the spectral propagator $G$ at a $k$-space grid point $(k_x,k_y) = (m\Delta k,n\Delta k)$ is averaged over the grid segment:
\begin{equation}
    \bar{G} = \frac{1}{(\Delta k)^2}
    \int_{(\Delta k)^2}G(k_x,k_y,z)dk_xdk_y
\end{equation}
Following Williams and Maynard, this integral can be evaluated in polar coordinates by approximating a square grid segment $(\Delta k)^2$ with a pie sector $k_0\Delta\phi\Delta k$:
\begin{multlinetwocolumn}\label{eq:polar-average}
    \bar{G} = \frac{1}{k_0\Delta\phi\Delta k}
    \int_{\phi_1}^{\phi_1+\Delta\phi}\int_{k_1}^{k_2}Gk_\rho dk_\rho = \twocolumnbreak
    \frac{2}{k_2^2 - k_1^2}\int_{k_1}^{k_2}Gk_\rho dk_\rho,
\end{multlinetwocolumn}
with $k_\rho^2 = k_x^2+k_y^2$, $k_1 = k_0 - \Delta k/2$, $k_2 = k_0 + \Delta k/2$, and $k_0^2 = (m^2 + n^2)\Delta k$.

Williams and Maynard only evaluate the integral for the case $d = 0$ for $G_\mathrm{vp}$. Here, we evaluate the integral for general $d$ for all three spectral propagators $G_\mathrm{vp}$, $G_\mathrm{pp}$, and $G_\mathrm{pv}$:
\begin{equation}
    \bar{G}_\mathrm{vp} = 
    \begin{cases}
        \displaystyle \frac{2}{k_2^2-k_1^2}\frac{k}{id}(
        e^{ik_{z,1}d} - 
        e^{ik_{z,2}d}),
        & \text{if } |d|\neq 0, k_\rho \neq 0 \\[10pt]
        \displaystyle \frac{2}{k_2^2-k_1^2}k(k_{z,1}-k_{z,2}),
        & \text{if } d = 0, k_\rho \neq 0 \\[10pt]
        e^{ikd},
        & \text{if } k_\rho = 0
    \end{cases}
\end{equation}

\begin{equation}
    \bar{G}_\mathrm{pp} = 
    \begin{cases}
        \displaystyle \frac{2}{k_2^2-k_1^2}\frac{1}{d^2}
        \left[(1 - idk_{z,1})e^{ik_{z,1}d}\right. \\
        \hfill - \left.(1 - idk_{z,2})e^{ik_{z,2}d}\right], \\
	\begin{array}{ll}
            & \text{if } |d|\neq 0, k_\rho \neq 0 \\
            1,
            & \text{if } d = 0\\
            e^{ikd},
            & \text{if } k_\rho = 0
	\end{array}
    \end{cases}
\end{equation}

\begin{equation}
    \bar{G}_\mathrm{pv} = 
    \begin{cases}
        \displaystyle \frac{2}{k_2^2-k_1^2}\frac{1}{kd^3}
        \left[(ik_{z,2}^2d^2 - 2k_{z,2}d - 2i)e^{ik_{z,2}d}\right. \\
        \hfill - \left.(ik_{z,1}^2d^2 - 2k_{z,1}d - 2i)e^{ik_{z,1}d}\right], \\
	\begin{array}{ll}
            & \text{if } |d|\neq 0, k_\rho \neq 0 \\
            \displaystyle \frac{2}{k_2^2-k_1^2}\frac{1}{3k}
            (k_{z,1}^3 - k_{z,2}^3),
            & \text{if } d = 0, k_\rho \neq 0 \\[6pt]
            e^{ikd},
            & \text{if } k_\rho = 0,
	\end{array}
    \end{cases}
\end{equation}
with $k_{z,1} = \sqrt{k^2 - k_1^2}$ and $k_{z,2} = \sqrt{k^2 - k_2^2}$. The cases $k_\rho = 0$ are not the outcome of integral~\ref{eq:polar-average}, but directly the value of $G(0,0,d)$, which is a good approximation of $\bar{G}(0,0,d)$ provided $\Delta k \ll k$.

\bibliographystyle{apsrev4-2}
\bibliography{references.bib}

\end{document}